\documentclass[10pt]{article}
\usepackage{amsmath}
\usepackage{amssymb}
\usepackage[dvips]{epsfig}

 \advance \textwidth by -2in
 \advance \textheight by -3in
\def\##1{\underline{#1}}
\def\=#1{\underline{\underline{#1}}}

\def\+#1{\underline{\bf #1}}
\def\*#1{\underline{\underline{\bf #1}}}

\def\r#1{(\ref{#1})}
\def\l#1{\label{#1}}
\def\c#1{\cite{#1}}

\def\le{\left(}
\def\ri{\right)}
\def\les{\left[}
\def\ris{\right]}
\def\lec{\left\{}
\def\ric{\right\}}

\def\.{\mbox{ \tiny{$^\bullet$} }}

\def\lambdao{\lambda_{\scriptscriptstyle 0}}

\def\eps{\epsilon}

\begin{document}

\begin{center}
\LARGE{{\bf On the effective permittivity of silver--insulator
nanocomposites }} \vskip 0.6cm

\normalsize

 {\bf \large{ Tom G. Mackay}} \\\vskip 0.4cm \normalsize
 School of Mathematics,
  University of Edinburgh,
 Edinburgh EH9 3JZ, UK \\ \vskip 0.2cm  {\sf T.Mackay@ed.ac.uk}
 \vskip 0.4cm

\end{center}

\vspace{15mm}

\noindent {\bf Abstract.} The Bruggeman formalism provides an
estimate of the effective permittivity of a composite material
comprising two constituent materials, with each constituent material
 being composed of electrically small particles.
 When one of the
constituent materials is silver and the other is an insulating
material, the Bruggeman estimate of the effective permittivity of
the composite exhibits resonances with respect to volume fraction
that are not physically plausible.

\vskip 0.6cm


\section{INTRODUCTION} \l{intro}

The estimation of the  electromagnetic properties of homogenized
composite materials (HCMs) is a fundamental problem which has
generated a vast literature, spanning 200 years \c{L96}. Interest in
this topic has escalated lately with the advent of
\emph{metamaterials} \c{Walser}; i.e., artificial composite
materials which exhibit properties either not exhibited at all by
their constituents or not exhibited to the same extent by their
constituents \c{M05}. In particular, metamaterials in the form of
nanocomposites~---~which are  assemblies of disparate
nanoparticles~---~present exciting possibilities.

An important category of HCM is considered in this communication:
that is, we consider HCMs which arise from the homogenization of two
isotropic dielectric constituent materials, namely materials $a$ and
$b$ with relative permittivities $\eps_a$ and $\eps_b$,
characterized by $\delta < 0$, where\footnote{The operators
$\mbox{Re} \lec \. \ric$ and $\mbox{Im} \lec \. \ric$ deliver the
real and imaginary parts of their complex--valued arguments.}
\begin{equation}
\delta = \frac{\mbox{Re} \lec \eps_a \ric }{\mbox{Re} \lec \eps_b
\ric}.
\end{equation}
A recent study demonstrated that conventional homogenization
formalisms, such as the Bruggeman and Maxwell Garnett formalisms, do
not necessarily provide physically plausible estimates of the HCM
relative permittivity in the $\delta < 0$ regime \c{ML04}.
Furthermore, much--used bounds on the HCM relative permittivity,
such as the Hashin--Shtrikman and Bergman--Milton bounds, can become
exceedingly large when $\delta < 0$ \c{DML06}.

Many metal--insulator HCMs of interest  belong to the $\delta < 0$
category, with silver often being the constituent metal of choice. A
key property of silver which is exploited by designers of
HCM--metamaterials is that its relative permittivity $\eps_{Ag}$ is
such that $\mbox{Re} \lec \eps_{Ag} \ric < 0$ but $ \mbox{Im} \lec
\eps_{Ag} \ric  \ll | \, \mbox{Re} \lec \eps_{Ag} \ric |$ at visible
and near infrared wavelengths, as illustrated in Figure~\ref{fig1}
by the plots of measured values of $\mbox{Re} \lec \eps_{Ag} \ric$
and $\mbox{Im} \lec \eps_{Ag} \ric$ provided in \c{Ag}. Whereas the
Bruggeman homogenization formalism has been widely applied to
estimate the relative permittivity of silver--insulator HCMs
\c{Granqvist,Roy,Aouadi,Velikov,deSande,Cai,Chettiar}, the inherent
limitations of the Bruggeman formalism in this particular $\delta <
0$ scenario are not widely appreciated. Herein we highlight the
possible pitfalls of applying the Bruggeman homogenization formalism
to silver--insulator HCMs.

 \setcounter{figure}{0}
\begin{figure}[!ht]
\centering \psfull \epsfig{file=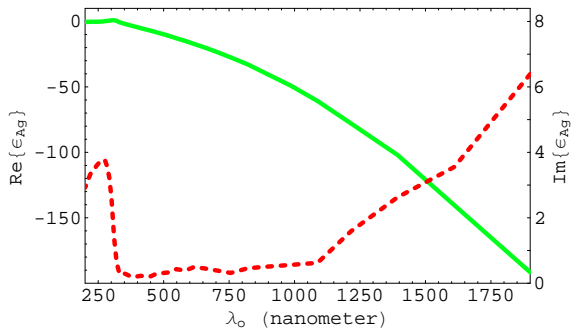,width=3.0in}
 \caption{ Real (green, solid curve) and imaginary (red, dashed curve)
  parts of the relative permittivity of silver as a function of wavelength (in nanometer). From tabulated values of experimental
  measurements provided
   in \c{Ag}.
  } \label{fig1}
\end{figure}

\section{NUMERICAL INVESTIGATIONS}

In the Bruggeman homogenization formalism, the constituent materials
are treated symmetrically. As a consequence, this approach may be
applied at arbitrary volume fractions, unlike the  Maxwell Garnett
homogenization formalism which is restricted to dilute composites. A
rigorous basis for the Bruggeman formalism is developed within the
framework of the strong--permittivity--fluctuation theory (SPFT)
\c{TK81, MLW00}.
 The Bruggeman estimate of the relative permittivity of the HCM,
arising from the homogenization of the two constituent materials
described in \S\ref{intro}, is provided implicitly as  \c{Ward}
\begin{eqnarray}
&& \eps_{Br} = \frac{f_a \eps_a \le \eps_b + 2 \eps_{Br} \ri + f_b
\eps_b \le \eps_a + 2 \eps_{Br} \ri }{f_a  \le \eps_b + 2 \eps_{Br}
\ri + f_b \le \eps_a + 2 \eps_{Br} \ri}, \l{Br}
\end{eqnarray}
wherein  $f_a$ and $f_b = 1 - f_a$ are the  respective volume
fractions of the constituent materials $a$ and $b$. The particles of
both constituent materials are assumed to be spherical and
electrically small. For example, at the (optical) wavelength of 600
nanometers,   constituent particles of less  than 60 nanometers in
radius are envisaged.

The solution
\begin{equation}
\eps_{Br} = \frac{1}{2} \lec -\les \eps_a \le f_b - 2f_a \ri +
\eps_b \le f_a - 2f_b \ri \ris \pm \sqrt{ \les \eps_a \le f_b - 2f_a
\ri + \eps_b \le f_a - 2f_b \ri \ris^2 + 8 \eps_a \eps_b}\ric
 \l{quadratic}
\end{equation}
is straightforwardly extracted from \r{Br}. The choice of sign for
the square root term in \r{quadratic} is dictated by the restriction
$\mbox{Im}\,\lec \eps_{Br} \ric \ge 0$, as per the Kramers--Kronig
relations \c{BH}.

\begin{figure}[!ht]
\centering \psfull \epsfig{file=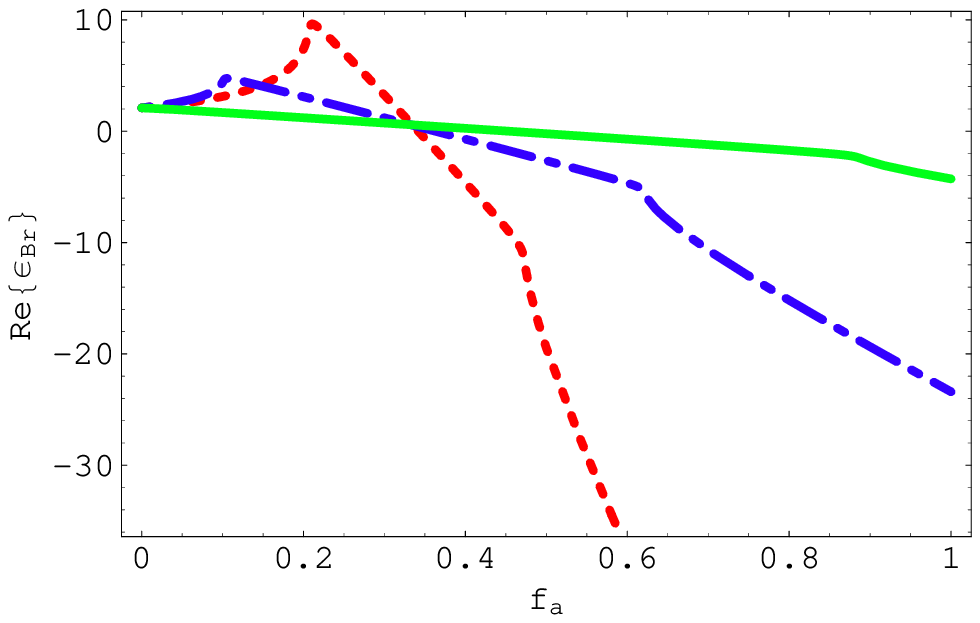,width=3.0in}
\vspace{5mm} \\ \epsfig{file=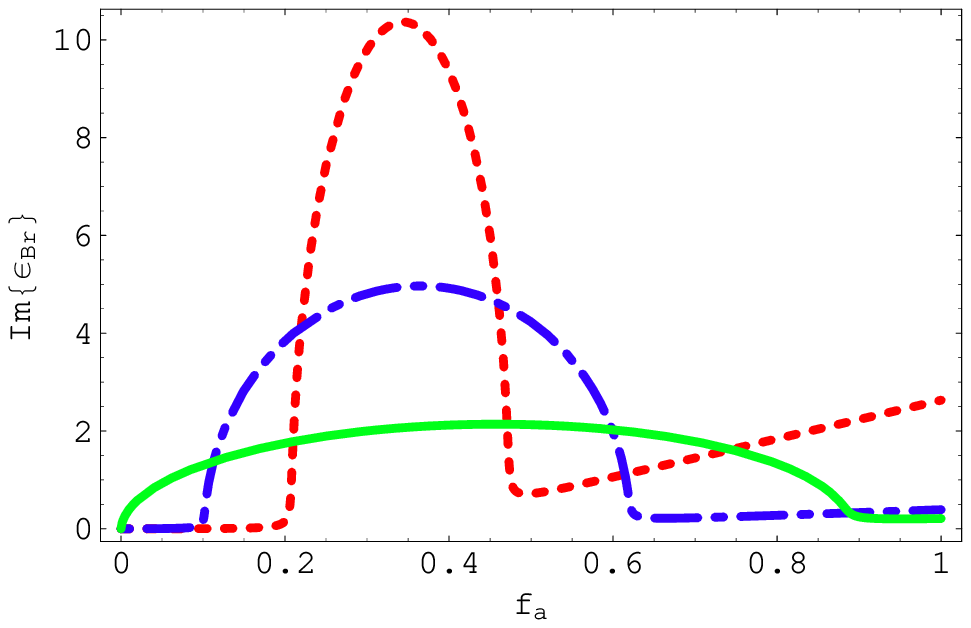,width=3.0in}
 \caption{The Bruggeman estimate $\eps_{Br}$ of the relative permittivity for
 the HCM arising from constituents with  $\eps_a =  \eps_{Ag}$ and $\eps_b = 2.1$. Real (top) and imaginary (bottom)
   parts are plotted against $f_a$ for $\lambdao = 1393$ nm (red, dashed curve), $\lambdao = 704$ nm (blue, broken dashed curve)
    and $\lambdao = 397$ nm (green, solid curve).
  } \label{fig2}
\end{figure}

Let $\eps_a = \eps_{Ag}$, as given by the values displayed in
Figure~\ref{fig1}. For constituent material $b$ we choose silica
with $\eps_b = 2.1$, in keeping with several reported
Bruggeman--based studies \c{Granqvist,Roy,Cai,Chettiar}. The real
and imaginary parts of the Bruggeman estimate $\eps_{Br}$ are
plotted as functions of volume fraction $f_a$ in Figure~\ref{fig2}.
At the wavelengths $\lambdao = 397, 704$ and $ 1393$ nanometers
considered in Figure~\ref{fig2}, the corresponding values of
$\eps_a$ are $-4.3 + 0.2i$, $-23.4 + 0.4i$ and $-102.0+2.6i$,
respectively \c{Ag}. The distinct resonances exhibited by $\mbox{Im}
\lec \eps_{Br} \ric$ with respect to volume fraction, and the
associated abrupt changes in gradient of $\mbox{Re} \lec \eps_{Br}
\ric$, are most striking. The $\mbox{Im} \lec \eps_{Br} \ric$
resonance range is $0 \lessapprox f_a \lessapprox 0.9$ for $\lambdao
= 397$ nm, $0.1 \lessapprox f_a \lessapprox 0.62$ for $\lambdao =
704$ nm and $0.2 \lessapprox f_a \lessapprox 0.47$ for $\lambdao =
1393$ nm. We observe that the maximum values of $\mbox{Im} \lec
\eps_{Br} \ric$ are an order of magnitude greater than the
 values of $\mbox{Im} \lec \eps_{a} \ric$ at the corresponding wavelengths.

\begin{figure}[!ht]
\centering \psfull \epsfig{file=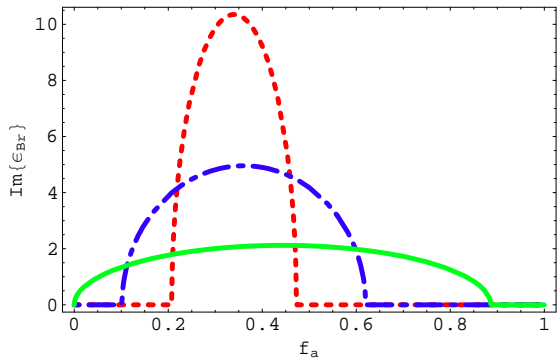,width=3.0in}
 \caption{The imaginary part of the Bruggeman estimate $\eps_{Br}$ of the  relative permittivity for the
  HCM arising from constituent materials with
   $\eps_a = \mbox{Re} \lec \eps_{Ag} \ric $ and $\eps_b = 2.1$,
    plotted against $f_a$ for $\lambdao = 1393$ nm (red, dashed curve), $\lambdao = 704$ nm (blue, broken dashed curve)
    and $\lambdao = 397$ nm (green, solid curve).
  } \label{fig3}
\end{figure}

The question arises: does the resonant behaviour illustrated in
Figure~\ref{fig2} represent a physical process? To address this
question, we repeat the calculations of Figure~\ref{fig2} using
$\mbox{Re} \lec \eps_{a} \ric = \mbox{Re} \lec \eps_{Ag} \ric$ but
$\mbox{Im} \lec \eps_{a} \ric = 0$. The corresponding Bruggeman
estimates $\mbox{Im} \lec \eps_{Br} \ric$ are graphed against volume
fraction $f_a$ in Figure~\ref{fig3}. It is  clear from
Figure~\ref{fig3} that the resonant behaviour illustrated in
Figure~\ref{fig2} persists in an almost identical manner even when
$\mbox{Im} \lec \eps_{a} \ric = 0$. In the case of Figure~\ref{fig3}
both constituent materials are nondissipative,  yet  the Bruggeman
estimate of the HCM relative permittivity corresponds to a HCM which
is strongly dissipative in the  regions of resonance. However, the
Bruggeman homogenization formalism has no mechanism for
accommodating coherent scattering losses, unlike the SPFT which is
its natural generalization \c{MLW00}. Therefore, we infer that the
Bruggeman estimates in the regions of resonance are not physically
plausible.

\section{CONCLUDING REMARKS}

The Bruggeman estimate of the relative permittivity for a
silver--insulator HCM~---~and, by extension, the SPFT estimate
\c{TK81, MLW00}~---~exhibits resonances with respect to volume
fraction which are not physically plausible. This point should be
carefully borne in mind when considering the effective permittivity
of silver--insulator nanocomposites. We note that in two recent
studies wherein the Bruggeman formalism was applied to estimate the
relative permittivity of silver--insulator HCMs, the resonance
region was excluded from consideration \c{Cai,Chettiar}.

\vspace{10mm}

\noindent {\bf Acknowledgement} TGM is supported by a \emph{Royal
Society of Edinburgh/Scottish Executive Support Research Fellowship}

\end{document}